\documentclass{article}

\usepackage{amsfonts}
\usepackage{bm}
\newcommand{\g}{\gamma}
\newcommand{\gt}{\tilde\gamma}
\newcommand{\bfg}{\boldsymbol{\gamma}}
\newcommand{\bfr}{\boldsymbol{\rho}}
\newcommand{\bfgt}{\boldsymbol{\tilde\gamma}}
\newcommand{\lm}{\lambda}
\newcommand{\lmt}{\tilde\lambda}
\newcommand{\bflm}{\boldsymbol{\lambda}}
\newcommand{\bfq}{\boldsymbol{q}}
\newcommand{\G}{\Gamma}
\newcommand{\ir}{{\rm i}}
\newcommand{\h}{\hbar}
\def\ba{\begin{array}}
\def\ea{\end{array}}
\def\bea{\begin{eqnarray}}
\def\eea{\end{eqnarray}}
\def\bea*{\begin{eqnarray*}}
\def\eea*{\end{eqnarray*}}
\def\be{\begin{equation}}
\def\ee{\end{equation}}
\def\A{{\cal A}}
\def\B{{\cal B}}
\def\C{{\cal C}}
\def\D{{\cal D}}

\begin{document}

\begin{center}
{\large
WAVE FUNCTIONS OF THE TODA CHAIN WITH BOUNDARY INTERACTION}
\end{center}

\bigskip

\begin{center}
Nikolai IORGOV, Vitaly SHADURA
\end{center}

\begin{center}
Bogolyubov Institute for Theoretical Physics
\end{center}

\begin{center}
03143, Kiev, Ukraine
\end{center}

\begin{abstract}
In this contribution, we give an integral representation of the
wave functions of the  quantum $N$-particle Toda chain with boundary
interaction. In the case of the Toda chain with one-boundary
interaction, we obtain the wave function by an integral
transformation from the wave functions of the open Toda chain. The
kernel of this transformation is given explicitly in terms of
$\Gamma$-functions. The wave function of the Toda chain with two-boundary
interaction is   obtained   from the previous wave functions
by  an integral transformation. In
this case, the difference equation for the kernel of the integral
transformation admits  separation of variables. The separated
difference equations coincide with  the Baxter equation.
\end{abstract}

\section{Introduction}

Recently, some progress in the derivation of the eigenfunctions of the
Hamiltonians of some integrable quantum chains with finite number
of  particles has been achieved \cite{Skl2}--\cite{KLS2}. It is
connected with the development of the method of separation of variables
\cite{Skl2} for quantum integrable models. The first steps in the
elaboration of this method were taken by Gutzwiller \cite{Gu}, who has
found a solution of the eigenvalue problem for $N=2,3,4$-particle
periodic Toda chain.

Using the $R$-matrix formalism,
Sklyanin \cite{Skl1} proposed an algebraic formulation of the
method of  separation  of variables  applicable to  a broader class of integrable
quantum chains.
The next important step was taken by Kharchev and
Lebedev \cite{KL1}, who combined the analytic method of Gutzwiller and algebraic approach
of Sklyanin. They obtained
the eigenfunctions of the  $N$-particle periodic Toda chain
by some integral transformation of the eigenfunctions of an auxiliary problem,
the  open ($N-1$)-particle
Toda chain. It turned out that the kernel of this transformation
admits  separation of variables. The separated equations coincide with the Baxter equation.
A solution of this equation  has been found in
\cite{PG} (see also \cite{KL1}).

Later Kharchev and Lebedev \cite{KL4} have found   a
remarkable recurrence relation between the eigenfunctions of
the $N$-particle and ($N-1$)-particle open Toda chains.
Understanding  these formulas from the  viewpoint of the
representation theory \cite{GKL1} made it   possible to extend their approach
to  other integrable systems \cite{GKL1,KLS2}.

In this paper, we apply this method to the  derivation of the
eigenfunctions of the commuting Hamiltonians of the $N$-particle
quantum Toda chain with boundary interaction. We use the Sklyanin
approach \cite{sklyanin:bnd} to the boundary problems for the
quantum integrable models. The $N$-particle
eigenfunctions of the quantum Toda chain in which the first and
last particles exponentially interact with  the walls (the two-boundary
interaction) is constructed by means of an integral
transformation of the eigenfunctions for the Toda chain with
one-boundary interaction (the auxiliary problem). These
eigenfunctions, in turn, are constructed using the
eigenfunctions of the  $N$-particle open Toda chain. Such a complicated
hierarchy allows one to separate the variables in the difference
equation for the kernel of the mentioned integral transformation
reducing it to a version  of the Baxter equation.
We note that, for the classical Toda chain with general boundary interaction,
the separation of variables was performed by Kuznetsov \cite{Kuzn}.

\section{Integrals of motion of the open Toda chain}

To describe the integrals of motion of the quantum $N$-particle open Toda chain, we use
the $L$-operators (one for each particle)
\[
L_k(u)=
\left( \begin{array}{lr}
u-p_k  & e^{-q_k}\\
-e^{q_k} & 0
\end{array}\right),\qquad k=1,2,\ldots,N,
\]
where $N$ is the number of particles in the chain,
$p_k$ and $q_k$ are the operators of momentum and position of the $k$-th particle,
respectively.
The monodromy matrix is defined as
\be\label{trmatr}
T(u):=L_N(u)L_{N-1}(u)\cdots L_2(u)L_1(u)=
\left( \begin{array}{lr}
A(u) & B(u)\\
C(u) & D(u)
\end{array}\right).
\ee
The commutation relations for the matrix elements of $T(u)$ follow
from the canonical commutation relations
\[
[p_k,q_l]=-\ir\h \delta_{kl}
\]
and can be written as
\be\label{RTT}
R(u-v)\left(T(u)\otimes {\bf 1}\right) \left({\bf 1} \otimes T(v) \right)=
\left( {\bf 1} \otimes T(v) \right) \left(T(u)\otimes {\bf 1}\right)R(u-v),
\ee
where $R(u)$ is the rational $R$-matrix:
\be\label{Rmatr}
R(u)=\left( \begin{array}{cccc}
1+\frac{\ir\h}{u} & 0 & 0 & 0 \\
0 & 1 & \frac{\ir\h}{u} & 0 \\
0 &  \frac{\ir\h}{u} & 1 & 0 \\
0 & 0 & 0 & 1+\frac{\ir\h}{u}
\end{array}\right).
\ee
From (\ref{trmatr}) it follows that $A(u)$ is a polynomial of degree $N$ in $u$:
\[
A(u)=\sum_{m=0}^N(-1)^m u^{N-m} H_m(p_1,q_1;p_2,q_2;\ldots;p_N,q_N)=
\]
\[
=u^N-H_1 u^{N-1}+H_2 u^{N-2}-\cdots +(-1)^N H_N.
\]
In particular, relations (\ref{RTT}) give
\[
[A(u),A(v)]=0,
\]
and, therefore, $[H_m,H_k]=0$, that is, $A(u)$ is a generating function for the
commuting operators $H_m$.
Since
\[
H_1=\sum_{k=1}^N p_k,\qquad
H_2=\sum_{k,l\atop k<l} p_k p_l-\sum_{k=1}^{N-1} e^{q_k-q_{k+1}},
\]
we get the Hamiltonian for the open Toda chain in the form
\[
H=H_1^2/2-H_2=\sum_{k=1}^N \frac{p_k^2}{2}+\sum_{k=1}^{N-1} e^{q_k-q_{k+1}}.
\]
Therefore, the operators $H_m$ are Hamiltonians for the open Toda chain.

\section{Wave functions for the open Toda chain}

Let a wave function $\psi(q_1,\ldots, q_N)$
for the open Toda chain be a common eigenfunction of the commuting Hamiltonians $H_m$:
\[
H_m \psi(q_1,\ldots, q_N)=E_m\psi(q_1,\ldots, q_N).
\]
Then
\[
A(u) \psi_{\bfg_N} (q_1,\ldots, q_N) =
\prod_{l=1}^N (u-\g_{Nl}) \psi_{\bfg_N} (q_1,\ldots, q_N),
\]
where $\bfg_N=(\g_{N1},\g_{N2},\ldots,\g_{NN})$
are the quantum numbers of the $N$-particle system,
$E_m=e_m(\g_{N1},\g_{N2},\ldots,\g_{NN})$, and $e_m$ is the $m$-th elementary
symmetric polynomial.
For every set $\bfg_N$, the space of eigenfunctions is $N!$ dimensional.
The physical eigenfunction $\psi_{\bfg_N}$ is fixed by the requirement that
$\psi_{\bfg_N}$ rapidly decreases in the classically forbidden region, that is, for
$q_k>>q_{k+1}$ for some $k$. For $q_1<<q_2<<\cdots<<q_N$, $\psi_{\bfg_N}$ is
a superposition of plane waves.

Recently, Kharchev and Lebedev \cite{KL4} have found a recursive procedure
of constructing the $N$-particle wave function $\psi_{\bfg_N}(q_1,q_2,\ldots,q_N)$
through the $(N-1)$-particle
wave functions $\psi_{\bfg_{N-1}}(q_1,q_2,\ldots,q_{N-1})$.
The recurrence relation is
\be\label{psiOT}
\psi_{\bfg_N}(q_1,q_2,\ldots,q_N)=\int d\bfg_{N-1} \mu(\bfg_{N-1})
Q(\bfg_{N-1}|\bfg_{N}) \psi_{\bfg_{N-1}}(q_1,q_2,\ldots,q_{N-1})
e^{\frac{\ir}{\h}(\sum_{j=1}^N \g_{N,j}-\sum_{k=1}^{N-1} \g_{N-1,k}) q_N},
\ee
where integration is carried out with respect to $\g_{N-1,k}$, $k=1,2,\ldots,N-1$,
along any set of the lines parallel to the real axis and such that
\be\label{ineqgg}
\min_k\, {\rm Im}\,\g_{N-1,k}>\max_j\,  {\rm Im}\,\g_{N,j},\qquad
 j=1,\ldots,N,
\ee
\be
Q({\bfg}_{N-1}|{\bfg}_N)=
\prod_{k=1}^{N-1}\prod_{j=1}^N \h^{\frac{\g_{N-1,k}-\g_{N,j}}{\ir\h}}
\G\left(\frac{\g_{N-1,k}-\g_{N,j}}{\ir\h}\right),
\qquad \mu^{-1}({\bfg_{N-1}})=
\prod_{k,l\atop k\ne l}
\G\left(\frac{\g_{N-1,k}-\g_{N-1,l}}{\ir\h}\right).
\ee
In a similar way, the $(N-1)$-particle wave functions can be expressed
through the $(N-2)$-particle wave functions,
and so on. The wave function for the $1$-particle open Toda chain is just a
plane wave:
\[
\psi_{\g_{11}}(q_1)=e^{\frac{\ir}{\h}\g_{11}q_1}.
\]

In what follows, we use the notation ${\bfg}:={\bfg}_N$, $\g_{k}:=\g_{N,k}$.
As shown in \cite{KL4}, the wave function $\psi_{\bfg}$ satisfies the relations
\be
A(u)\psi_{\bfg}=\prod_{l=1}^N(u-\gamma_l)\,
\psi_{\bfg},
\label{ap}
\ee
\be
B(u)\psi_{\bfg}={\rm i}^{N-1}
\sum_{p=1}^N
\left(\prod_{l\ne p}\frac{u-\gamma_l}{\gamma_p-\gamma_l}\right)\,
\psi_{\bfg^{+p}},
\label{bp}
\ee
\be
C(u)\psi_{\bfg}={\rm i}^{-N-1} \sum_{p=1}^N
\left(\prod_{l\ne p}\frac{u-\gamma_l}{\gamma_p-\gamma_l}\right)\,
\psi_{\bfg^{-p}},
\label{cp}
\ee
where $\psi_{\bfg^{\pm p}}:= \psi_{\g_1,\g_2,
\ldots,\g_p\pm {\rm i} \hbar,\ldots, \g_N }$.

In order to find the action of $D(u)$ on $\psi_{\bfg}$,
we use the following property of the
quantum determinant of $T(u)$ for the Toda chain:
\be\label{qdet}
D(u)A(u-{\rm i} \hbar)-C(u)B(u-{\rm i} \hbar)=1.
\ee
The result is
\[
D(u)\psi_{\bfg}=\sum_{p=1}^N
\left(\prod_{l\ne p}\frac{u-\gamma_l}{\gamma_p-\gamma_l}\right)
\frac 1{{\rm i}\hbar}
\left(\frac 1{\prod_{l\ne p}(\gamma_p-\gamma_l+{\rm i}\hbar)}-
\frac 1{\prod_{l\ne p}(\gamma_p-\gamma_l-{\rm i}\hbar)}\right)
 \psi_{\bfg}-
\]
\be
-\sum_{p,q\atop p\ne q}\frac 1 {\gamma_q-\gamma_p-{\rm i}\hbar}
\frac 1{\prod_{l\ne p} (\gamma_p-\gamma_l)}
\prod_{l\ne p,q}\frac{u-\gamma_l}{\gamma_q-\gamma_l}
\ \psi_{\bfg^{+p,-q}}\,.
\label{dp}
\ee

\section{Integrals of motion of the Toda chain with boundary interaction}

In this section, we give a sketch of the $R$-matrix formalism
for the quantum Toda chain with boundary
interaction proposed by Sklyanin \cite{sklyanin:bnd}.
This formalism is important for the construction of wave functions.
The key object in this approach is the matrix
\be
U(u):=T(u)K^{(-)}(u-\frac {\ir\h}2){\tilde T(-u)}=\left( \begin{array}{lr}
{\cal A}(u) & {\cal B}(u)\\
{\cal C}(u) & {\cal D}(u)
\end{array}\right),
\label {tru}
\ee
where $T(u)$ is the monodromy matrix (\ref{trmatr}) of
the $N$-particle open Toda chain, and
\[
{\tilde T(-u)}=\sigma_2 {T^t(-u)} \sigma_2=(\sigma_2 L^t_1(-u)\sigma_2)
(\sigma_2 L^t_2(-u)\sigma_2)\cdots (\sigma_2 L^t_N(-u)\sigma_2).
\]
Here, $\sigma_2$ is the Pauli matrix.
The matrix $K^{(-)}(u-\ir\h/2)$ is
\be
K^{(-)}(u-\frac {\ir\h}2)= \left( \begin{array}{lr}
\alpha_1&u-\frac {\ir\h}2
\\
-\beta_1(u-\frac {\ir\h}2) & \alpha_1
\end{array}\right).
\ee
As shown in \cite{sklyanin:bnd}, the matrix $U(u)$ satisfies
the reflection equation
\be
R(u-v)\left(U(u)\otimes {\bf 1}\right) R(u+v-\ir\h) \left({\bf 1}\otimes U(v)\right)=
\left({\bf 1}\otimes U(v)\right)
R(u+v-\ir\h)\left(U(u)\otimes {\bf 1}\right) R(u-v),
\label{RE}
\ee
where $R(u)$ is given by (\ref{Rmatr}).

This equation implies
${\cal B}(u){\cal B}(v)={\cal B}(v){\cal B}(u)$. Therefore,
the expansion of ${\cal B}(u)$ in powers of
$u$ gives commuting operators which, in fact, are
the Hamiltonians of the one-boundary Toda chain
\be \label{exprB}
{\cal B}(u)=(-1)^N (u-\ir\h/2)
\left(u^{2N}-u^{2N-2}H_1^{\rm B}+ u^{2N-4} H_2^{\rm B}-\cdots+(-1)^N H_N^{\rm B}\right),
\ee
where
\[
H_1^{\rm B}=\sum_{k=1}^N p_k^2+2\sum_{k=1}^{N-1}e^{q_k-q_{k+1}}
-2\alpha_1 e^{-q_1}+\beta_1 e^{-2q_1}.
\]
Here the last two terms describe interaction of the first particle with the wall.

The Sklyanin's transfer-matrix
\be
t(u):={\rm Tr}\, K^{(+)}(u+\ir\h/2)U(u),
\label{montr}
\ee
where
\[
K^{(+)}(u+\frac {\ir\h}2)= \left( \begin{array}{lr}
\alpha_N&\beta_N(u+\frac {\ir\h}2)\\
-(u+\frac {\ir\h}2) & \alpha_N
\end{array}\right),
\]
satisfies the commutation relation \cite{sklyanin:bnd}
\be
t(u)t(v)=t(v)t(u).
\label{tcom}
\ee
Hence, $t(u)$ is a generating function for commuting
operators which, in fact, are the Hamiltonians of the two-boundary Toda chain.

For simplicity, in what follows, we fix $\beta_1=\beta_N=0$ and
use the notation $e^{\kappa_1}:=-2\alpha_1$, $e^{-\kappa_N}:=-2\alpha_N$.
In this case, we have
\be\label{tH}
t(u)=(-1)^{N-1} (u^2+\h^2/4)
\left(u^{2N}-u^{2N-2}H_1^{\rm BB}+ u^{2N-4} H_2^{\rm BB}-\cdots+(-1)^N H_N^{\rm BB}\right)
+2\alpha_1\alpha_N,
\ee
where
\[
H_1^{\rm BB}=\sum_{k=1}^N p_k^2+2\sum_{k=1}^{N-1}e^{q_k-q_{k+1}}-2\alpha_1 e^{-q_1}-2\alpha_N e^{q_N}.
\]

In the case of the Toda chain, the matrix $U(u)$ has some additional
symmetry (unitarity) \cite{sklyanin:bnd}:
\be\label{unitarity}
\left( \begin{array}{cc}
{\cal A}(-u) & {\cal B}(-u)\\
{\cal C}(-u) & {\cal D}(-u)
\end{array}\right)=\frac{1}{2u-\ir\h}
\left( \begin{array}{cc}
-\ir\h \A(u)+2u\D(u) & -(2u+\ir\h)\B(u)\\
-(2u+\ir\h)\C(u) & 2u \A(u)-{\ir\h} \D(u)
\end{array}\right).
\ee
In particular, this leads to
\be\label{AD}
{\cal A} (u)=\frac{1}{u}
\left((u-\frac{\ir\h}{2}){\cal D}(-u)+\frac{\ir\h}{2}{\cal D}(u)\right).
\ee
Therefore, using this equality and (\ref{montr}), we obtain
\be\label{tu}
t(u)= \alpha_N\frac{(u+\frac{\ir\h}{2})}{u}\ {\cal D} (u)+
\alpha_N\frac{(u-\frac{\ir\h}{2})}{u}\ {\cal D} (-u)
-(u+\frac{\ir\h}{2}){\cal B}(u).
\ee

Using (\ref{tru}), we obtain the following expressions for the matrix elements of $U(u)$
in terms of the
matrix elements of the monodromy matrix $T(u)$ for the $N$-particle open Toda chain:
\be
 {\cal A}(u)=\alpha_1\left(A(u)D(-u)-B(u)C(-u)\right)-
 \left(u-\ir\frac \h 2\right)A(u)C(-u),
\label{a}
\ee
\be
{\cal B}(u)=-\alpha_1\left(A(u)B(-u)-B(u)A(-u)\right)+
\left(u-\ir\frac \h 2\right)A(u)A(-u),
\label{b}
\ee
\be
 \C(u)=\alpha_1 \left(C(u)D(-u)-D(u)C(-u)\right)
 -\left(u-\ir\frac \h 2\right)C(u)C(-u),
\label{c}
\ee
\be
{\cal D}(u)=\alpha_1\left( D(u)A(-u)-C(u)B(-u)\right)+
\left(u-\ir\frac \h 2\right)C(u)A(-u).
\label{d}
\ee

We give some examples:
\[
\underline{N=1:} \qquad \B(u)=-(u-\ir\h/2) (u^2-(p_1^2+e^{\kappa-q_1})),
\]
\[
t(u)=(u^2+\h^2/4)\left(u^2-(p^2+e^{\kappa-q_1}+e^{q_1-\kappa'})\right)+2\alpha \alpha';
\]
\[
\underline{N=2:} \qquad \B(u)=(u-\ir\h/2)
\left(u^4-u^2(p_1^2+p_2^2+2e^{q_1-q_2}
+e^{\kappa_1-q_1})+(p_1 p_2-e^{q_1-q_2})^2
-\alpha_1 p_2^2e^{-q_1}-2\alpha_1 e^{-q_2}\right),
\]
\[
t(u)=-(u^2+\h^2/4)\left(u^4-u^2(p_1^2+p_2^2+2e^{q_1-q_2}
+e^{\kappa_1-q_1}+e^{q_2-\kappa_2})+\cdots \right)+2\alpha_1 \alpha_2.
\]

\section{Wave functions for the one-boundary Toda chain}

We define the function $\Psi_{\bflm}\equiv
\Psi_{\lambda_1, \ldots,\lambda_N}$ as
\be \label{psilm}
\Psi_{\bflm}(q_1,\ldots,q_N)=\int d\gamma_1\cdots d\gamma_N
\mu({\bfg}) Q({\bfg}|{\bflm})
e^{-\frac{\ir \kappa_1 (\g_1+\cdots+\g_N)}{\h}}\psi_{\bfg}(q_1,\ldots,q_N),
\ee
where $e^{\kappa_1}=-2\alpha_1$ and
\be \label{muQ}
Q({\bfg}|{\bflm})=
\frac{\prod_{k,l}\G\left(\frac{\lm_l-\g_k}{\ir\h}\right)\G
\left(\frac{-\lm_l-\g_k}{\ir\h}\right)}
{\prod_{k,l \atop k<l}\G\left(-\frac{\g_k+\g_l}{\ir\h}\right)}
\prod_k\h^{-\frac{(N+1)\g_k}{\ir\h}},
\qquad \mu^{-1}({\bfg})=\prod_{k,l\atop k\ne
l}\G\left(\frac{\g_k-\g_l}{\ir\h}\right).
\ee
We show that this is a wave function for the quantum one-boundary
Toda chain, and
\be
\B(u)\Psi_{\bflm}(q_1,\ldots,q_N)=
(-1)^N (u-\frac{\ir\h}2)\prod_{l=1}^N{(u^2-\lm_l^2)}
\Psi_{\bflm}(q_1,\ldots,q_N),
\label{bpl}
\ee
where the structure of the right-hand side corresponds to (\ref{exprB}).
The integration in (\ref{psilm}) is carried out
along any set of lines parallel to the real axis and such that
\be\label{ineqlmgm}
\max_k\, {\rm Im}\,\g_{k}< -\min_j\, {\rm Im}\,\lm_j,
\qquad k=1,2,\ldots,N,\quad j=1,\ldots,N.
\ee

First, we prove the absolute convergence in (\ref{psilm}). For this,  we
use the inequalities
\[
|\Gamma(x+\ir y)|\le \Gamma(x) p_x(|y|) e^{-\frac{\pi |y|}{2}}, \qquad x>0,
\]
where $p_x(|y|)$ is some polynomial in $|y|$ with degree linearly depending on $x$,
\[
\frac{1}{|\Gamma(x+\ir y)|}\le \frac{\left(1+\frac{|y|}{x}\right)
e^{\frac{\pi|y|}{2}}}{\Gamma(x)},
\qquad x>0,
\]
and also (\ref{mainineq}) from Appendix~A:
\[
\sum_{k,l=1}^N \left(|\tilde\lm_k-\tilde\g_{N,l}|+|\tilde\lm_k+\tilde\g_{N,l}|\right)
+\sum_{r=1}^{N-1}\sum_{k,l} |\tilde\g_{r+1,k}-\tilde\g_{r,l}|
-2\sum_{r=2}^{N}\sum_{k<l} |\tilde\g_{r,k}-\tilde\g_{r,l}|
-\sum_{k<l} |\tilde\g_{N,k}+\tilde\g_{N,l}|
\ge
\]
\[
\ge
-2 N \sum_{k=1}^N|\tilde\lm_k|+
2 \sum_{k<l} \left(|\tilde\lm_k-\tilde\lm_l|+|\tilde\lm_k+\tilde\lm_l|\right)
+\frac{2}{N}\sum_{r=1}^N \sum_{k=1}^r |\tilde\g_{r,k}|,
\]
which is valid for any set of real variables $\tilde\lm_k$, $k=1,2,\ldots,N$;
$\tilde\g_{r,l}$, $l=1,2,\ldots,r$, $r=1,2,\ldots,N$.
A proof of the last inequality is given in Appendix~A.
For our purposes, we fix $\tilde\lm_k$ (respectively, $\tilde\g_{r,l}$) to be equal
to ${\rm Re}\,\lm_k$ (respectively, ${\rm Re}\,\g_{r,l}$).

Presenting (\ref{psilm}) as
\[
\Psi_{\bflm}(q_1,\ldots,q_N)=\int\prod_{r=1}^N \prod_{k=1}^r d\tilde\g_{r,k}
F(\bfg_1,\bfg_2,\ldots,\bfg_N,\bflm;q_1,\ldots,q_N),
\]
we obtain the following inequality for the dependence of the integrand on $\g_{r,k}$:
\be\label{estim}
|F(\bfg_1,\bfg_2,\ldots,\bfg_N,\bflm;q_1,\ldots,q_N)|\le
P(\{\tilde\g_{r,k}\})
\exp\left(-\frac{\pi}{\h N}\sum_{r=1}^N\sum_{k=1}^r |\tilde\g_{r,k}|\right),
\ee
where $P(\{\tilde\g_{r,k}\})$ has polynomial dependence
on the variables $\tilde\g_{r,k}$ and certain dependence on the other variables.
Estimate (\ref{estim}) leads to absolute convergence of the integral on
the right-hand side of (\ref{psilm}).
We would like to mention that integral (\ref{psilm}) does not depend on the values
of the imaginary parts of $\g_{r,k}$ (that is, lines of integration)
provided the mentioned inequalities (\ref{ineqgg}) and (\ref{ineqlmgm})
for them are satisfied.
This follows from two facts.
First, we do not encounter poles as we shift the integration contour.
Second, due to estimate (\ref{estim}), the integrand is vanishing at the
infinities of the integration contours.
This justifies the correctness of shifting of the integration contours which
we use in what follows.

From the physical viewpoint,
the function $\Psi_{\bflm}(q_1,\ldots,q_N)$ given by (\ref{psilm}) has correct
asymptotic behaviour rapidly decreasing in the classically forbidden region, that is,
where $q_k>>q_{k+1}$ for some $k$ or where $q_1<<0$.
In the region $0<<q_1<<q_2<<\cdots<<q_N$,
the function $\Psi_{\bflm}(q_1,\ldots,q_N)$ is
a superposition of plane waves.

The formulas for the action of the matrix elements of $U(u)$ on $\Psi_{\bflm}$,
in particular (\ref{bpl}), are derived in Appendix~B.

Here we give some heuristic explanation of formulas (\ref{apadp})
\[
{\cal D}(\lm_r) \Psi_{\bflm}= \alpha_1 \Psi_{\bflm^{-r}}, \qquad
{\cal D}(-\lm_r) \Psi_{\bflm} = \alpha_1 \Psi_{\bflm^{+r}},
\]
which are proved in Appendix~B.
Let $\Psi_{\bflm}(q_1,\ldots, q_N)$ be an
eigenfunction of $\B(u)$ satisfying (\ref{bpl}).
Then the commutation relation
\[
\left(u^2-(v-\ir\h)^2\right)\D(v)\B(u)-(u^2-v^2)\B(u)\D(v)=
\ir\h(u+v-\ir\h)\D(u)\B(v)+\ir\h (u-v)\A(u)\B(v),
\]
which follows from (\ref{RE}), gives
\[
\B(u)\D(\lm_r)\Psi_{\bflm}
=(-1)^N (u-\frac{\ir\h}2)  (u^2-(\lm_r-\ir\h)^2) \prod_{k=1\atop k\ne r}^{N}(u^2-\lm_k^2)
\cdot\D(\lm_r)\Psi_{\bflm}
\]
at $v=\lm_r$, and, therefore, $\D(\lm_r)\Psi_{\bflm}$ is an eigenfunction of
${\cal B}(u)$ with $\lm_r$ replaced by $(\lm_r-\ir\h)$.  Clearly, this argumentation is
not sufficient to prove the relation
${\cal D}(\lm_r) \Psi_{\bflm}= \alpha_1 \Psi_{\bflm^{-r}}$.

\section{Wave functions for the two-boundary Toda chain}

Taking into account (\ref{tH}), it is useful to introduce
\[
\tilde t(u) :=(-1)^{N-1} \frac{t(u)-2\alpha_1\alpha_N}{u^2+\left(\frac
{\h}{2}\right)^2}=
u^{2N}-u^{2N-2}H_1^{\rm BB}+ u^{2N-4} H_2^{\rm BB}-\cdots+(-1)^N H_N^{\rm BB}.
\]

Let $\Phi_{\bfr}(\bfq)$ be a wave function for the two-boundary Toda chain:
\[
\tilde t(u) \Phi_{\bfr}(\bfq) = \prod_{k=1}^N (u^2-\rho_k^2) \Phi_{\bfr}(\bfq)
=: \tilde t (u|\bfr) \Phi_{\bfr}(\bfq),
\]
where $\bfr=\{\rho_1,\rho_2,\ldots,\rho_N\}$ are the quantum numbers of
the corresponding state.

We look for  $\Phi_{\bfr}(\bfq)$ in the form
\be\label{intphi}
\Phi_{\bfr}(\bfq)= \int d\lm_1\cdots d\lm_N\
\tilde\mu({\bflm})C({\bflm}|{\bfr})\Psi_{ {\bflm}}(\bfq),
\ee
where
\[
\tilde\mu^{-1}(\bflm)= \prod_{i,j\atop i < j}
\left(\Gamma\left(\frac{\lm_i-\lm_j}{\ir\h}\right)
\Gamma\left(-\frac{\lm_i-\lm_j}{\ir\h}\right)\right)
\prod_{i,j\atop i \le j}
\left(\Gamma\left(\frac{\lm_i+\lm_j}{\ir\h}\right)
\Gamma\left(-\frac{\lm_i+\lm_j}{\ir\h}\right)\right),
\]
and the integration with respect to $\{\lm_k\}$ is carried out along arbitrary
lines parallel to the real axis.
Using (\ref{apat}) and
\[
\frac{\tilde\mu(\bflm^{+p})}{\tilde\mu(\bflm)}=\frac{(\lm_p+\ir\h)}{\lm_p}\prod_{l\ne p}
\frac{(\lm_p+\ir\h)^2-\lm_l^2}{\lm_p^2-\lm_l^2},
\]
we obtain
\[
(-1)^{N-1} \tilde t(u|\bfr)\Phi_{\bfr}(\bfq)=\int d\lm_1\cdots d\lm_N\
\Psi_{{\bflm}}(\bfq)
\left[ \alpha_1\alpha_N \sum_{p=1}^{N}\left[
\frac{\tilde\mu(\bflm^{+p})C(\bflm^{+p}|\bfr)}{(\lm_p+\ir\h)(\lm_p+\frac{\ir\h}{2})}
 \left(\prod_{l\ne p}\frac{u^2-\lm_l^2}{(\lm_p+\ir\h)^2-\lm_l^2}\right)+\right.\right.
\]
\[
+\left.\left.\frac{\tilde\mu(\bflm^{-p})C(\bflm^{-p}|\bfr)}{(\lm_p-\ir\h)(\lm_p-\frac{\ir\h}{2})}
\left(\prod_{l\ne p}\frac{u^2-\lm_l^2}{(\lm_p-\ir\h)^2-\lm_l^2}\right) -
\frac{2\tilde\mu(\bflm)C(\bflm |\bfr)}{(\lm_p^2+\frac{\h^2}{4})}
\left(\prod_{l\ne p}\frac{u^2-\lm_l^2}{\lm_p^2-\lm_l^2}\right)
\right]-\prod_{l=1}^N{(\lm_l^2-u^2)}\right]=
\]
\[
= \int d\lm_1d\lm_2\cdots d\lm_N\
\mu({\bflm})C({\bflm}|\bfr)\Psi_{ {\bflm}}(\bfq) \left[ \alpha_1\alpha_N
\sum_{p=1}^{N}\left(\prod_{l\ne p}
\frac{u^2-\lm_l^2}{(\lm_p+\ir\h)^2-\lm_l^2}\right)\right.\times
\]
\[\times
\left. \left[ \frac{1}{\lm_p(\lm_p+\frac{\ir\h}{2})}\frac{
C(\bflm^{+p}|\bfr)}{C(\bflm|\bfr)}+
\frac{1}{\lm_p(\lm_p-\frac{\ir\h}{2})}\frac{
C(\bflm^{-p}|\bfr)}{C(\bflm|\bfr)}-\frac{2}{(\lm_p^2+\frac{\h^2}{4})}\right]-
\prod_{l=1}^N (\lm_l^2-u^2)\right].
\]
We set  $u=\lm_p$. Then the previous relation is satisfied if
\[
(-1)^{N-1}\tilde t(\lm_p|\bfr)=\frac{t(\lm_p|\bfr)-2\alpha_1\alpha_N}{\lm_p^2+
\frac{\h^2}{4}}=
\]
\[
=\alpha_1\alpha_N\left[
\frac{1}{\lm_p(\lm_p+\frac{\ir\h}{2})}\frac{
C(\bflm^{+p}|\bfr)}{C(\bflm|\bfr)}+
\frac{1}{\lm_p(\lm_p-\frac{\ir\h}{2})}\frac{
C(\bflm^{+p}|\bfr)}{C(\bflm|\bfr)}-\frac{2}{(\lm_p^2+\frac{\h^2}{4})}\right],
\]
where $t(u|\bfr)=(-1)^{N-1}(u^2+\h^2/4)\prod_{k=1}^N (u^2-\rho_k^2)+2\alpha_1\alpha_N$.
This multidimensional difference equation admits separation of variables.
Namely, we suppose the factorization property
\[
C(\bflm|\bfr)=\prod_{p=1}^N\ c(\lm_p|\bfr).
\]
Then $c(\lm|\bfr)$ satisfies the Baxter equation
\[
\frac{1}{\lm(\lm+\frac{\ir\h}{2})} c(\lm+\ir\h|\bfr)+
\frac{1}{\lm(\lm-\frac{\ir\h}{2})}
c(\lm-\ir\h|\bfr)=\frac{t(\lm|\bfr)c(\lm|\bfr)}{\alpha_1\alpha_N(\lm^2+\frac{\h^2}{4})},
\]
or, equivalently,
\[
(\lm-\frac{\ir\h}{2}) c(\lm+\ir\h|\bfr)+ (\lm+\frac{\ir\h}{2})
c(\lm-\ir\h|\bfr)=\frac{\lm \ t(\lm|\bfr)\ c(\lm|\bfr)}{\alpha_1\alpha_N}.
\]

Solutions of this equation can be constructed in terms of ratios of
 infinite-dimensional determinants as it was done in the case
of the periodic Toda chain \cite{PG,KL1}.
We expect that, similarly to the case of the
periodic Toda chain \cite{PG,KL1}, the requirement of the analytical properties
of $c(\lm|\bfr)$ (which is  important,
in particularly, for the convergence of  integral (\ref{intphi})) restricts possible
 values of $\bfr$ to the discrete spectrum
of the quantum two-boundary Toda chain.

\section*{Acknowledgment}
The authors are grateful to S.~Kharchev, D.~Lebedev and S.~Pakuliak
for valuable stimulating discussions,
 and to the organizers of the
conference ``Classical and Quantum Integrable Systems''
for their support, which made it possible
to present the results of this work at the conference.
 The research  was partially supported by the INTAS Grant No. 03-51-3350 and by
the State Foundation for Basic Research of Ukraine, Project
2.7/00152.

\section*{Appendix A: A proof of some inequalities}

\noindent
{\bf Proposition.} {\it
For any  two sets of {\bf real} variables
$\bfgt_{N-1}=$ $\{\gt_{N-1,1},$ $\gt_{N-1,2},\ldots,$ $\gt_{N-1,N-1}\}$
and $\bfgt_{N}=$ $\{\gt_{N,1},$ $\gt_{N,2},\ldots,$ $\gt_{N,N}\}$,
the following inequality is valid:
\begin{equation} \label{FIneq}
F(\bfgt_{N-1},\bfgt_N):=\sum_{k=1}^N \sum_{l=1}^{N-1} |\gt_{N,k}-\gt_{N-1,l}|-
\sum_{k_1,k_2=1 \atop k_1< k_2}^N |\gt_{N,k_1}-\gt_{N,k_2}|-
\sum_{l_1,l_2=1 \atop l_1< l_2}^{N-1} |\gt_{N-1,l_1}-\gt_{N-1,l_2}|\ge 0.
\end{equation}
}

\noindent
{\bf Proof.}
Since $F(\bfgt_{N-1},\bfgt_N)$ is invariant with respect to permutations
of elements in the sets $\bfgt_{N-1}$
and  $\bfgt_{N}$, we can restrict ourselves to the case where
$\gt_{N-1,1}\ge$ $\gt_{N-1,2}\ge \cdots$ $\ge\gt_{N-1,N-1}$ and
$\gt_{N,1}\ge$ $\gt_{N,2}\ge \cdots$ $\ge\gt_{N,N}$.
In this case, simple combinatorics shows that
\[
\sum_{k_1< k_2}^N |\gt_{N,k_1}-\gt_{N,k_2}|+
\sum_{l_1< l_2}^{N-1} |\gt_{N-1,l_1}-\gt_{N-1,l_2}|
=\sum_{k=1}^N \left(\sum_{l=1}^{k-1}(\gt_{N-1,l}-\gt_{N,k})+\sum_{l=k}^{N-1}(\gt_{N,k}-\gt_{N-1,l}) \right).
\]
In fact, both sides are equal to
\[
\sum_{k=1}^N (N+1-2k)\gt_{N,k}+\sum_{l=1}^{N-1} (N-2k)\gt_{N-1,l}.
\]
Since $|a|-a\ge 0$ for any real $a$, we have
\[
F(\bfgt_{N-1},\bfgt_N)=\sum_{k=1}^N \left(
\sum_{l=1}^{k-1}\left(|\gt_{N-1,l}-\gt_{N,k}|-(\gt_{N-1,l}-\gt_{N,k})\right)
+\sum_{l=k}^{N-1}\left(|\gt_{N,k}-\gt_{N-1,l}|-(\gt_{N,k}-\gt_{N-1,l})\right) \right)\ge 0.
\]
We  note that $F(\bfgt_{N-1},\bfgt_N)=0$ if and only if
\[
\gt_{N,1}\ge \gt_{N-1,1} \ge \gt_{N,2}\ge \gt_{N-1,2}\ge
\cdots \ge \gt_{N,N-1} \ge \gt_{N-1,N-1} \ge \gt_{N,N}.\hspace{50mm} \Box
\]

As a corollary, we obtain
\[
G(\bfgt_{N-1},\bfgt_N):=F(\bfgt_{N},\{\gt_{N-1,1},\gt_{N-1,2},\ldots,\gt_{N-1,N-1}, 0 ,0 \})
=\sum_{k=1}^N \sum_{l=1}^{N-1} |\gt_{N,k}-\gt_{N-1,l}|-
\]
\begin{equation} \label{GIneq}
-\sum_{k_1< k_2}^N |\gt_{N,k_1}-\gt_{N,k_2}|-
\sum_{l_1< l_2}^{N-1} |\gt_{N-1,l_1}-\gt_{N-1,l_2}|
+2\sum_{k=1}^N |\gt_{N,k}|-2\sum_{l=1}^{N-1} |\gt_{N-1,l}|\ge 0.
\end{equation}

For $N$ arbitrary sets $\bfgt_{s}=$ $\{\gt_{s,1},$ $\gt_{s,2},\ldots,$ $\gt_{s,s}\},$
$s=1,\ldots,N$,  of {\bf real} numbers, combining  inequalities
(\ref{FIneq}) and (\ref{GIneq}), we get
\[
H_s:=\sum_{r=1}^{s-1} F(\bfgt_{r},\bfgt_{r+1})
+\sum_{r=s}^{N-1} G(\bfgt_{r},\bfgt_{r+1})\ge 0,
\qquad s=1,\ldots,N.
\]
Explicitly,
\[
H_s:=\sum_{r=1}^{N-1}
\sum_{k=1}^{r+1} \sum_{l=1}^{r} |\gt_{r+1,k}-\gt_{r,l}|
-\sum_{k<l}^N |\gt_{N,k}-\gt_{N,l}|
-2 \sum_{r=2}^{N-1} \sum_{k<l}^r |\gt_{r,k}-\gt_{r,l}|+
\]\[
+2\sum_{k=1}^N |\gt_{N,k}|-2\sum_{l=1}^{s} |\gt_{s,l}|\ge 0,
\qquad s=1,\ldots,N.
\]
Therefore, the inequality $\sum_{s=1}^N H_s/N\ge 0$ is equivalent to
\[
\sum_{r=1}^{N-1} \sum_{k=1}^{r+1} \sum_{l=1}^{r} |\gt_{r+1,k}-\gt_{r,l}|
-\sum_{k<l}^N |\gt_{N,k}-\gt_{N,l}|
-2 \sum_{r=2}^{N-1} \sum_{k<l}^r |\gt_{r,k}-\gt_{r,l}|\ge
\]
\begin{equation} \label{ineqsh}
\ge -2\sum_{k=1}^N |\gt_{N,k}|+\frac 2 N \sum_{s=1}^N \sum_{l=1}^{s} |\gt_{s,l}|.
\end{equation}

We need one more inequality for real $\lmt_1,$ $\lmt_2,\ldots,$ $\lmt_N$ and
$\gt_{N,1},$ $\gt_{N,2},\ldots,$ $\gt_{N,N}$.
It is just  inequality (\ref{GIneq}) for
$G(\{\lmt_1,$ $\lmt_2,\ldots,$ $\lmt_N,$ $-\lmt_1,$ $-\lmt_2,\ldots,$ $-\lmt_N\},$
$\{\gt_{N,1},$ $\gt_{N,2},\ldots,$ $\gt_{N,N},$ $0,0,\ldots,$ $0\})$
with $N-1$ zeros. It is
\[
\sum_{k,l=1}^N \left(|\tilde\lm_k-\tilde\g_{N,l}|+|\tilde\lm_k+\tilde\g_{N,l}|\right)
- \sum_{k<l}^N \left(|\tilde\g_{N,k}-\tilde\g_{N,l}|+|\tilde\g_{N,k}+\tilde\g_{N,l}|\right)
\ge
\]
\be\label{ineqlmg}
\ge -2 N \sum_{k=1}^N|\tilde\lm_k|+
2 \sum_{k<l}^N \left(|\tilde\lm_k-\tilde\lm_l|+|\tilde\lm_k+\tilde\lm_l|\right)+
2\sum_{k=1}^N|\tilde\g_{N,k}|.
\ee
Adding (\ref{ineqsh}) and (\ref{ineqlmg}),  we obtain the main inequality
\[
\sum_{k,l=1}^N \left(|\tilde\lm_k-\tilde\g_{N,l}|+|\tilde\lm_k+\tilde\g_{N,l}|\right)
+\sum_{r=1}^{N-1}\sum_{k,l} |\tilde\g_{r+1,k}-\tilde\g_{r,l}|
-2\sum_{r=2}^{N}\sum_{k<l} |\tilde\g_{r,k}-\tilde\g_{r,l}|
-\sum_{k<l} |\tilde\g_{N,k}+\tilde\g_{N,l}|
\ge
\]
\be\label{mainineq}
\ge
-2 N \sum_{k=1}^N|\tilde\lm_k|+
2 \sum_{k<l} \left(|\tilde\lm_k-\tilde\lm_l|+|\tilde\lm_k+\tilde\lm_l|\right)
+\frac{2}{N}\sum_{r=1}^N \sum_{k=1}^r |\tilde\g_{r,k}|.
\ee

\section*{Appendix B: Formulas  for action of the matrix elements
of $U(u)$ on $\Psi_{\bflm}$}

In this Appendix, we  prove the following action formulas:
\be
\B(u)\Psi_{\bflm}=
(-1)^N (u-\frac{\ir\h}2)\prod_{l=1}^N{(u^2-\lm_l^2)}
\Psi_{\bflm},
\label{apbpl}
\ee
\[
 {\cal D}(u)\Psi_{\bflm}=\alpha_1 \sum_{p=1}^N \left(
\prod_{l\ne p}\frac{u^2-\lm_l^2}{\lm_p^2-\lm_l^2}\right)
\left[\frac{(u+\lm_p)}{2\lm_p} \frac{(u-\frac {\ir\h}{2})}{(\lm_p-
\frac {\ir\h}{2})}\Psi_{\bflm^{-p}}+ \frac{(u-\lm_p)}{2\lm_p}
\frac{(u-\frac {\ir\h}{2})}{(\lm_p+ \frac
{\ir\h}{2})}\Psi_{\bflm^{+p}}\right]+
\]
\be
+ \alpha_1
\left(\prod_{l=1}^N\frac{\lm_l^2-u^2}{\lm_l^2+(\frac
{\h}{2})^2}\right)\Psi_{\bflm},
\label{apdpsi}
\ee
\[
\tilde t(u)\Psi_{\bflm}= (-1)^{N-1}
\frac{t(u)-2\alpha_1\alpha_N}{u^2+\left(\frac
{\h}{2}\right)^2}\Psi_{\bflm}=\prod_{l=1}^N{(u^2-\lm_l^2)}\Psi_{\bflm} +
\]
\be\label{apat}
+ (-1)^{N-1}\alpha_1 \alpha_N \sum_{p=1}^N
 \left(\prod_{l\ne p}\frac{u^2-\lm_l^2}{\lm_p^2-\lm_l^2}\right)
 \left[
\frac{1}{\lm_p(\lm_p- \frac {\ir\h}{2})}\Psi_{\bflm^{-p}}\ +
\frac{1}{\lm_p(\lm_p+ \frac {\ir\h}{2})} \Psi_{\bflm^{+p}}-
\frac{2}{{\lm_p^2+\left(\frac{\h}{2}\right)^2}}\Psi_{\bflm}\right].
\ee
In particular,  formula (\ref{apdpsi}) gives
\be\label{apadp}
{\cal D}(\lm_r) \Psi_{\bflm}= \alpha_1 \Psi_{\bflm^{-r}}, \qquad
 {\cal D}(-\lm_r) \Psi_{\bflm} = \alpha_1 \Psi_{\bflm^{+r}},\qquad
{\cal D}({\ir\h}/{2})\Psi_{\bflm}  =
\alpha_1\Psi_{\bflm}.
\ee
The action of ${\cal A}(u)$ and ${\cal C}(u)$
on $\Psi_{\bflm}$  can be derived using (\ref{AD}) and
Sklyanin determinant \cite{sklyanin:bnd}
for $U(u)$, respectively.

Before presenting a proof of the  action formulas, we give  useful relations for
$\mu(\bfg)$ and $Q({\bfg}|{\bflm})$ from (\ref{muQ}):
\be\label{mu1}
\frac{\mu(\bfg^{+q})}{\mu(\bfg)}=
\prod_{l\ne q}\frac{\g_l-\g_q-\ir\h}{\g_q-\g_l},\qquad
\frac{\mu(\bfg^{-p})}{\mu(\bfg)}=
\prod_{l\ne p}\frac{\g_p-\g_l-\ir\h}{\g_l-\g_p},
\ee
\be
\frac{\mu(\bfg^{-p,+q})}{\mu(\bfg)}=
\frac{\g_p-\g_q-2\ir\h}{\g_p-\g_q}
\prod_{l\ne p,q}\frac{\g_p-\g_l-\ir\h}{\g_l-\g_p}
\cdot\frac{\g_l-\g_q-\ir\h}{\g_q-\g_l},
\label{mu2}
\ee
\be \frac{Q(\bfg^{+q}|{\bflm})}{Q({\bfg}|{\bflm})}=
-\ir^{N+1}\frac{\prod_{l\ne q}(\g_q+\g_l+\ir\h)
}{\prod_l(\lm_l^2-(\g_q+\ir\h)^2)},\
\qquad \frac{Q(\bfg^{-p}|{\bflm})}{Q({\bfg}|{\bflm})}=
\ir^{-N+1}\frac{\prod_l(\lm_l^2-\g_p^2)}
{\prod_{l\ne p}(\g_p+\g_l)}, \label{Q1}
\ee
\be\label{Q2}
\frac{Q(\bfg^{-p,+q}|{\bflm})}{Q({\bfg}|{\bflm})}=
\prod_l\frac{\g_p^2-\lm_l^2}{(\g_q+\ir\h)^2-\lm_l^2}
\cdot\prod_{l\ne p,q}\frac{\g_l+\g_q+\ir\h}{\g_l+\g_p}.
\ee

\medskip
\noindent
{\it Action formula for} ${\cal B}(u)$.
\medskip

To prove formula (\ref{apbpl}), we use (\ref{b}). Taking into account
(\ref{ap}) and (\ref{bp}), we get
\[
\left(A(u)B(-u)-B(u)A(-u)\right)\Psi_{\bflm}=
2\,{\rm i}^{N-1}(u-\frac{\ir\h}2)
\sum_{p=1}^N
\int d\gamma_1\cdots d\gamma_N
\mu({\bfg}) Q({\bfg}|{\bflm})\times
\]
\[
\times e^{-\frac{\ir \kappa_1 (\g_1+\cdots+\g_N)}{\h}}
 \left(\prod_{l\ne p}\frac{ \gamma_l^2-u^2}{\gamma_p-\gamma_l}
 \right)\
  \psi_{\bfg^{+p}}.
\]
In every summand with fixed $p$,
we make the change of variable $\g_p\to\g_p-\ir\h$.
Then we use formulas (\ref{mu1}) and (\ref{Q1})
to transform $\mu({\bfg})$ and $Q({\bfg}|{\bflm})$
to the original forms. This leads to an additional factor.
Finally, we shift the contour of integration with respect to $\g_p$ to the
original one. This possibility was  explained in Section~5.
Thus, we have
\[
-\alpha_1\left(A(u)B(-u)-B(u)A(-u)\right)\Psi_{\bflm}=
\]
\[
=-2\alpha_1 {\rm i}^{N-1}(u-\frac{\ir\h}2)e^{- \kappa_1}
 \sum_{p=1}^N \int
d\gamma_1\cdots d\gamma_N \mu({\bfg^{-p}})
Q({\bfg^{-p}}|{\bflm})
e^{-\frac{\ir \kappa_1 (\g_1+\cdots+\g_N)}{\h}}
 \left(\prod_{l\ne p}\frac{ \gamma_l^2-u^2}{\gamma_p-\gamma_l
 -{\rm i}\hbar}
 \right)\
  \psi_{\bfg}=
\]
\[
=(u-\frac{\ir\h}2)
 \sum_{p=1}^N
 \int
d\gamma_1\cdots d\gamma_N \mu({\bfg}) Q({\bfg}|{\bflm})
e^{-\frac{\ir \kappa_1 (\g_1+\cdots+\g_N)}{\h}}
\left(\prod_{l\ne p}\frac{\gamma_l^2-u^2}{\g_l^2-\g_p^2}
\right) \prod_l(\lm_l^2-\g_p^2)
 \,  \psi_{\bfg},
\]
where we used $e^{\kappa_1}=-2\alpha_1$. Since
\[
(u-\frac{\ir\h}2)A(u)A(-u)\Psi_{\bflm}=
(u-\frac{\ir\h}2)
 \int
d\gamma_1\cdots d\gamma_N \mu({\bfg}) Q({\bfg}|{\bflm})
e^{-\frac{\ir \kappa_1
(\g_1+\cdots+\g_N)}{\h}}\prod_{l}{(\gamma_l^2-u^2)}
 \
  \psi_{\bfg},
\]
we get
\[
\B(u)\Psi_{\bflm}= (u-\frac{\ir\h}2)
 \int
d\gamma_1\cdots d\gamma_N \mu({\bfg}) Q({\bfg}|{\bflm})
e^{-\frac{\ir \kappa_1 (\g_1+\cdots+\g_N)}{\h}}\times
\]
\be
 \times\left[ \sum_{p=1}^N\left(\prod_{l\ne p}
 \frac{\gamma_l^2-u^2}{\g_l^2-\g_p^2} \right)
\prod_l(\lm_l^2-\g_p^2)+\prod_{l}{(\gamma_l^2-u^2)}\right]
 \
  \psi_{\bfg}.
\label{actb}
\ee

We prove the relation
\be
\sum_{p=1}^N\left(\prod_{l\ne p}
\frac{\gamma_l^2-u^2}{\g_l^2-\g_p^2} \right)
\prod_{l=1}^N(\lm_l^2-\g_p^2)+\prod_{l=1}^N{(\gamma_l^2-u^2)}
 =\prod_{l=1}^N{(\lm_l^2-u^2)}.
\label{pol}
\ee
It follows from the identity
\begin{equation}\label{id1}
\prod_{l=1}^N{(b_l-x)}-\prod_{l=1}^N{(a_l-x)}=
\sum_{p=1}^N\left(\prod_{l\ne p}
\frac{a_l-x}{a_l-a_p} \right)
\prod_{l=1}^N(b_l-a_p)
\end{equation}
for the variables $x$, $a_l$, $b_l$, $l=1,2,\ldots,n$,
if one substitutes $x:=u^2$, $a_l:=\gamma_l^2$,
$b_l:=\lambda_l^2$. Identity (\ref{id1}) is just the Lagrange
interpolation formula for the polynomial
$\prod_{l=1}^N{(b_l-x)}-\prod_{l=1}^N{(a_l-x)}$ in $x$ of degree $N-1$
reconstructed from its values at $N$ points $x=a_p$,
$p=1,2,\ldots,N$. Using (\ref{actb}) and (\ref{pol}), we obtain
(\ref{apbpl}).

\medskip
\noindent
{\it Action formula for} ${\cal D}(u)$.
\medskip

Our strategy is to derive (\ref{apadp}) and
then to reconstruct formula (\ref{apdpsi}) using the
Lagrange interpolation formula.
Having in mind the expression (\ref{d}) and the action
formulas (\ref{ap})--(\ref{cp}) and (\ref{dp}), we begin calculation.
We have
\[
C(u)B(-u)\psi_{\bfg}=
-\sum_{p,q \atop p\ne q} \frac{(u-\gamma_p-\ir\h)}{(\g_q-\g_p-\ir\h)}
\frac{(u+\gamma_q)}{(\g_q-\g_p)} \left(\prod_{l\ne p,q}
\frac{\gamma_l^2-u^2}{(\g_l-\g_p)(\g_l-\g_q)}\right)
 \psi_{\bfg^{+p,-q}}
\]
\[
-\sum_{p} \left(\prod_{l\ne p} \frac{\gamma_l+u}{\g_l-\g_p}\right)
\left(\prod_{l\ne p} \frac{u-\gamma_l}{\g_p-\g_l+\ir\h}\right) \
\psi_{\bfg},
\]
\[
D(u)A(-u)\psi_{\bfg}=-  \sum_{p,q\atop p\ne q}
\frac{(u+\gamma_p)}{(\g_q-\g_p-\ir\h)}
\frac{(u+\gamma_q)}{(\g_p-\g_q)}
 \left(\prod_{l\ne p,q}
\frac{\gamma_l^2-u^2}{(\g_l-\g_p)(\g_l-\g_q)}\right)
 \psi_{\bfg^{+p,-q}}+
\]
\[+\sum_{p} \left(\prod_{l\ne
p}\frac{\gamma_l^2-u^2}{\gamma_p-\gamma_l}\right) \frac 1{{\rm
i}\hbar} \left(\frac{-(u+\g_p)} {\prod_{l\ne
p}(\gamma_p-\gamma_l+{\rm i}\hbar)}- \frac{-(u+\g_p)}
{\prod_{l\ne p}(\gamma_p-\gamma_l-{\rm i}\hbar)}\right) \psi_{\bfg}.
\]

Therefore,
\[
 (D(u)A(-u)-C(u)B(-u))\psi_{\bfg}=
\]
\[
-2(u-\frac{\ir\h}2)\sum_{p,q\atop p\ne q}
\frac{u+\g_q}{(\g_p-\g_q)(\g_q-\g_p-\ir\h)}
\prod_{l\ne p,q}
\frac{(\g_l^2-u^2)}{(\g_p-\g_l)(\g_q-\g_l)}
\cdot\psi_{\bfg^{+p,-q}} +
\]
\[
+ \sum_p \left(\prod_{l\ne p} \frac{\g_l^2-u^2}{\g_p-\g_l}\right)
\left(\prod_{l\ne
p}\frac{1}{(\g_p-\g_l+\ir\h)}-\frac{(u+\g_p)}{\ir\h} \prod_{l\ne
p}\frac 1{(\g_p-\g_l+\ir\h)}+\frac{(u+\g_p)}{\ir\h}
\prod_{l\ne p}\frac 1{(\g_p-\g_l-\ir\h)}\right)\psi_{\bfg}.
\]

After the action of $D(u)A(-u)-C(u)B(-u)$ on $\Psi_{\bflm}$, the
integrand becomes a linear combination of $\psi_{\bfg}$ with
non-shifted indices $\bfg$ and $\psi_{{\bfg}^{+p,-q}}$ with all
possible $p$, $q$, $p\ne q$.
Changing variables and shifting the integration contours
as was described above for ${\cal B}(u)$, we
rewrite the result as
\[
\int d\gamma_1\cdots d\gamma_N \mu({\bfg})
Q({\bfg}|{\bflm})\left(\sum_{p,q\atop p\ne q}R_{p,q}(u) \right)
e^{-\frac{\ir(\g_1+\cdots+\g_N)\kappa_1}{\h}}\psi_{\bfg},
\]
where
\[
R_{p,q}(u)=-2(u-\frac{\ir\h}2)\frac{\mu(\bfg^{-p,+q})}{\mu(\bfg)}
\frac{Q(\bfg^{-p,+q}|{\bflm})}{Q({\bfg}|{\bflm})}
\frac{(u+\g_q+\ir\h)}{(\g_p-\g_q-2\ir\h)(\g_q-\g_p+\ir\h)}
\prod_{l\ne
p,q}\frac{(\g_l^2-u^2)}{(\g_p-\g_l-\ir\h)(\g_q-\g_l+\ir\h)}=
\]
\[
=2\left(u-\frac{\ir\h}2\right) \frac{(u+\g_q+\ir\h)}{(
\g_q-\g_p)(\g_q-\g_p+\ir\h)} \left(\prod_{l\ne p,q}
\frac{(\g_l^2-u^2)}{(\g_p-\g_l)(\g_q-\g_l)}
\frac{\g_l+\g_q+\ir\h}{\g_l+\g_p}\right)
\left(\prod_l\frac{\g_p^2-\lm_l^2}{(\g_q+\ir\h)^2-\lm_l^2}
\right).
\]

Thus, we have
\[
(D(u)A(-u)-C(u)B(-u))\Psi_{\bflm}=
\]
\[
=\int d\gamma_1\cdots d\gamma_N \mu({\bfg})
Q({\bfg}|{\bflm})\left(\sum_{p,q\atop p\ne q} R_{p,q}(u) + \sum_q
\left( R_q^{(1)}(u)+ R_q^{(2)}(u)+ R_q^{(3)}(u)\right)\right)
e^{-\frac{\ir(\g_1+\cdots+\g_N)\kappa_1}{\h}}\psi_{\bfg},
\]
where
\[
R_q^{(1)}(u)= \prod_{l\ne q}\frac{(\g_l^2-u^2)}{(\g_q-\g_l)(\g_q-\g_l+\ir\h)},
\]
\[
R_q^{(2)}(u)=-\frac{(u+\g_q)}{\ir\h}\cdot
\prod_{l\ne q}\frac{(\g_l^2-u^2)}{(\g_q-\g_l)(\g_q-\g_l+\ir\h)},
\]
\[
R_q^{(3)}(u)=\frac{(u+\g_q)}{\ir\h}\cdot
\prod_{l\ne q}\frac{(\g_l^2-u^2)}{(\g_q-\g_l)(\g_q-\g_l-\ir\h)}.
\]
Since
\[
C(u)A(-u)\psi_{\bfg}=
-\ir^{-N-1}\sum_p(u+\g_p) \left(\prod_{l\ne
p}\frac{\gamma_l^2-u^2}{\gamma_p-\gamma_l}\right)
\psi_{{\bfg}^{-p}},
\]
after appropriate shift
of the integration contours and change of variables,
the action of $(u-\ir\h/2) C(u)A(-u)$ on $\Psi_{\bflm}$ becomes
\[
(u-\ir\h/2) C(u)A(-u)\Psi_{\bflm}=\alpha_1 \int d\gamma_1\cdots
d\gamma_N \mu({\bfg})
Q({\bfg}|{\bflm})\left(\sum_{q}R_{q}(u)\right)
e^{-\frac{\ir(\g_1+\g_2+\cdots+\g_N)\kappa_1}{\h}}\psi_{\bfg},
\]
where
\[
R_q(u)=2(u-\frac{\ir\h}2)\frac{\mu(\bfg^{+q})}{\mu(\bfg)}
\frac{Q(\bfg^{+q}|{\bflm})}{Q({\bfg}|{\bflm})}
\ir^{-N-1}(u+\g_q+\ir\h) \prod_{l\ne q}
\frac{\gamma_l^2-u^2}{\gamma_q-\gamma_l+\ir\h}=
\]
\[
=-2(u-\frac{\ir\h}2)(u+\g_q+\ir\h)
\prod_{l\ne q}\frac{(u^2-\g_l^2)(\g_q+\g_l+\ir\h)}{(\g_q-\g_l)}
\prod_l\frac{1}{\lm_l^2-(\g_q+\ir\h)^2 }.
\]
We denote
\[
R_q^{(0)}(u):=R_q(u)+\sum_{p\ne q} R_{p,q}(u)=
\]
\[
 =-2(u-\frac{\ir\h}2)(u+\g_q+\ir\h) \prod_{l\ne
q}\frac{(u^2-\g_l^2)(\g_q+\g_l+\ir\h)}{(\g_q-\g_l)}
\cdot\prod_l\frac{1}{\lm_l^2-(\g_q+\ir\h)^2 }+
\]
\be\label{rq0}
+2\left(u-\frac{\ir\h}2\right) \sum_{p\ne q}
  \frac{(u+\g_q+\ir\h)} {(
\g_q-\g_p)(\g_q-\g_p+\ir\h)} \left(\prod_{l\ne p,q}
\frac{(\g_l^2-u^2)}{(\g_p-\g_l)(\g_q-\g_l)}
\frac{\g_l+\g_q+\ir\h}{\g_l+\g_p}\right)\cdot
\prod_l\frac{\g_p^2-\lm_l^2}{(\g_q+\ir\h)^2-\lm_l^2}.
\ee

Now we  prove the following relation:
\begin{equation}\label{relR0}
R_q^{(0)}(\lm_r)=\frac{2(\lm_r-\ir\h/2)}{(\g_q-\lm_r+\ir\h)}
\prod_{l\ne q}\frac{\g_l^2-\lm_r^2}{(\g_q-\g_l)(\g_q-\g_l+\ir\h)}.
\end{equation}
We start from the  identity
for the variables $a_1,a_2,\ldots,a_N$ and $b_1,b_2,\ldots,b_{N-1}$:
\[
\sum_{p=1}^N\frac{\prod_{l=1}^{N-1}(a_p-b_l)}{\prod_{l\ne p}(a_p-a_l)}=1.
\]
Separating the summand with $p=q$, we rewrite the identity in the form
\[
1+\sum_{p\ne q} \left(\prod_{l\ne p,q}\frac 1{a_p-a_l}\right)
\frac 1{a_q-a_p}\prod_{l=1}^{N-1}(a_p-b_l)
=\frac{\prod_{l=1}^{N-1} (a_q-b_l)}{\prod_{l\ne q} (a_q-a_l)}.
\]
Dividing this equality by its right-hand side, we get
\[
\frac{\prod_{l\ne q} (a_q-a_l)}{\prod_{l=1}^{N-1} (a_q-b_l)}+
\sum_{p\ne q} \left(\prod_{l\ne p,q}\frac{a_q-a_l}{a_p-a_l}\cdot
\prod_{l=1}^{N-1}\frac{a_p-b_l}{a_q-b_l}\right)=1.
\]
With $a_q:=(\g_q+\ir\h)^2$, $a_l:=\g_l^2$
($l\ne q$),
$\{b_1,\ldots,$ $b_{N-1}\}:=$
$\{\lm_1^2,\ldots,$
$\lm_{r-1}^2,$ $\lm_{r+1}^2,\ldots,$ $\lm_N^2\}$, we have
\[
\frac{\prod_{l\ne q} ((\g_q+\ir\h)^2-\g_l^2)}{\prod_{l\ne r}
((\g_q+\ir\h)^2-\lm_l^2)}+ \sum_{p\ne q} \left(\prod_{l\ne
p,q}\frac{(\g_q+\ir\h)^2-\g_l^2}{\g_p^2-\g_l^2}\cdot
\prod_{l\ne r}\frac{\g_p^2-\lm_l^2}{(\g_q+\ir\h)^2-\lm_l^2}\right) =1.
\]
Multiplying the obtained identity by
\[
\frac{2(\lm_r-\ir\h/2)}{(\g_q-\lm_r+\ir\h)} \prod_{l\ne
q}\frac{(\g_l^2-\lm_r^2)}{(\g_q-\g_l)(\g_q-\g_l+\ir\h)},
\]
after simple transformations, we obtain (\ref{relR0}).
Using this result, we get
\[
R_q^{(0)}(\lm_r)+R_q^{(1)}(\lm_r)+R_q^{(2)}(\lm_r)
= -\frac{1}{\ir\h(\g_q-\lm_r+\ir\h)}
\frac{\prod_{l}(\g_l^2-\lm_r^2)}
{\prod_{l\ne q}(\g_q-\g_l)(\g_q-\g_l+\ir\h)}.
\]

Consider the identity
\[
\sum_{m=1}^{2N}\prod_{s=1 \atop s \ne m}^{2N} \frac{u-a_s}{a_m-a_s}=1
\]
with $\{a_1,\ldots,$ $a_{2N}\}:=$
$\{\g_1,\ldots,$ $\g_{N},$ $\g_1+\ir\h,\ldots,$ $\g_N+\ir\h\}$ and $u=\lm_r$.
Explicitly, we have
\[
\sum_{q=1}^N\left\{\frac{1}{\ir\h}\cdot
\frac{\prod_{l\ne q}(\g_l-\lm_r)\prod_{l}(\g_l-\lm_r+\ir\h)}
{\prod_{l\ne q}(\g_q-\g_l)(\g_q-\g_l-\ir\h)}-
\frac{1}{\ir\h}\cdot
\frac{\prod_{l}(\g_l-\lm_r)\prod_{l\ne q}(\g_l-\lm_r+\ir\h)}
{\prod_{l\ne q}(\g_q-\g_l)(\g_q-\g_l+\ir\h)}
\right\}=1.
\]
Multiplying the both sides by
${\prod_{l}(\g_l+\lm_r)}/{\prod_{l}(\g_l-\lm_r+\ir\h)}$,
we get
\[
\sum_{q=1}^N\left\{
\frac{(\g_q+\lm_r)}{\ir\h}\cdot
\prod_{l\ne q}\frac{(\g_l^2-\lm_r^2)}{(\g_q-\g_l)(\g_q-\g_l-\ir\h)}
-\frac{1}{\ir\h(\g_q-\lm_r+\ir\h)} \cdot
\frac{\prod_{l}(\g_l^2-\lm_r^2)}{\prod_{l\ne q}(\g_q-\g_l)(\g_q-\g_l+\ir\h)}
\right\}=
\]\[=\frac{\prod_{l}(\g_l+\lm_r)}{\prod_{l}(\g_l-\lm_r+\ir\h)},
\]
which, in fact, coincides with
\[
\sum_{q=1}^N \left(R_q^{(3)}(\lm_r)+\left( R_q^{(0)}
(\lm_r)+R_q^{(1)}(\lm_r)+R_q^{(2)}(\lm_r)
\right)\right)=
\frac{\prod_{l}(\g_l+\lm_r)}{\prod_{l}(\g_l-\lm_r+\ir\h)}.
\]
Since
\[
Q(\bfg|\bflm)\frac{\prod_{l}(\g_l+\lm_r)}{\prod_{l}(\g_l-\lm_r+\ir\h)}=
Q(\bfg|{\bflm^{-r}}),
\]
we obtain ${\cal D}(\lm_r) \Psi_{\bflm}= \alpha_1 \Psi_{\bflm^{-r}}$.
In complete analogy with the above derivation, we find
${\cal D}(-\lm_r) \Psi_{\bflm}= \alpha_1 \Psi_{\bflm^{+r}}$.
From (\ref{d}) and (\ref{qdet}) at $u={\ir\h}/{2}$, it follows that
${\cal D}({\ir\h}/{2}) = \alpha_1$. Thus, we proved (\ref{apadp}).

From (\ref{ap}) (resp. (\ref{bp}), (\ref{cp}), (\ref{dp})), it follows that
the polynomial
${A}(u)$ (resp. ${B}(u)$, ${C}(u)$, ${D}(u)$)
has degree $N$ (resp. $(N-1)$, $(N-1)$, $(N-2)$) in $u$.
Hence, using (\ref{d}),
we find that polynomial ${\cal D}(u)$ has degree $2N$ in $u$.
We know the results (\ref{apadp}) of the action of ${\cal D}(u)$ on
$\Psi_{\bflm}$ in $2N+1$ points.
Applying the Lagrange interpolation formula, we obtain (\ref{apdpsi}).

\medskip
\noindent
{\it Action formula for} $t(u)$.
\medskip

Now we calculate the action of $t(u)$, given by (\ref{tu}), on $\Psi_{\bflm}$.
Using formulas (\ref{apbpl}) and (\ref{apdpsi}), we find
\[
t(u)\Psi_{\bflm}
=\left[u^2+\left(\frac {\h}{2}\right)^2\right]\left[
 \alpha_1 \alpha_N \sum_{p=1}^N  \left(\prod_{l\ne
p}\frac{u^2-\lm_l^2}{\lm_p^2-\lm_l^2}\right) \left(
\frac{1}{\lm_p(\lm_p- \frac {\ir\h}{2})}\Psi_{\bflm^{-p}}\ +
\frac{1}{\lm_p(\lm_p+ \frac {\ir\h}{2})}
\Psi_{\bflm^{+p}}\right)\right.+
\]
\[
 +\left. \frac{2\alpha_1\alpha_N}{u^2+\left(\frac {\h}{2}\right)^2}
\left(\prod_{l=1}^N\frac{\lm_l^2-u^2}{\lm_l^2+(\frac
{\h}{2})^2}\right)\Psi_{\bflm}
   -\prod_{l=1}^N{(\lm_l^2-u^2)}\Psi_{\bflm}\right].
\]
The identity
\[
\frac{1}{u^2+\left(\frac{\h}{2}\right)^2}
\left[\left(\prod_{l=1}^N\frac{\lm_l^2-u^2}{\lm_l^2+(\frac
{\h}{2})^2}\right) -1\right] = -\sum_{p=1}^N
\frac{1}{\lm_p^2+\left(\frac {\h}{2}\right)^2}
\left(\prod_{l\ne p}\frac{u^2-\lm_l^2}{\lm_p^2-\lm_l^2}\right),
\]
which can be obtained using the Lagrange interpolation formula,
implies (\ref{apat}).

\end{document}